\documentclass[12pt,preprint]{aastex}
\usepackage{textcomp}
\usepackage{amsmath}
\usepackage{graphicx}
\usepackage[T1]{fontenc}
\usepackage{rotating}

\def\beq{\begin{eqnarray}}
\def\eeq{\end{eqnarray}}

\begin{document}

\title{Jet Luminosity of Gamma-ray Bursts: Blandford-Znajek Mechanism v.s. Neutrino Annihilation Process}

\author{Tong Liu\altaffilmark{1,2,3,4}, Shu-Jin Hou\altaffilmark{5,2}, Li Xue\altaffilmark{1,4}, and Wei-Min Gu\altaffilmark{1,4}}

\altaffiltext{1}{Department of Astronomy and Institute of Theoretical Physics and Astrophysics, Xiamen University, Xiamen, Fujian 361005, China; tongliu@xmu.edu.cn}
\altaffiltext{2}{Key Laboratory for the Structure and Evolution of Celestial Objects, Chinese Academy of Sciences, Kunming, Yunnan 650011, China}
\altaffiltext{3}{State Key Laboratory of Theoretical Physics, Institute of Theoretical Physics, Chinese Academy of Sciences, Beijing 100190, China}
\altaffiltext{4}{SHAO-XMU Joint Center for Astrophysics, Xiamen University, Xiamen, Fujian 361005, China}
\altaffiltext{5}{College of Physics and Electronic Engineering, Nanyang Normal University, Nanyang, Henan 473061, China}

\begin{abstract}
A neutrino-dominated accretion flow (NDAF) around a rotating stellar-mass black hole (BH) is one of the plausible candidates for the central engine of gamma-ray bursts (GRBs). Two mechanisms, i.e., Blandford-Znajek (BZ) mechanism and neutrino annihilation process, are generally considered to power GRBs. Using the analytic solutions from \citet{Xue2013} and ignoring the effects of the magnetic field configuration, we estimate the BZ and neutrino annihilation luminosities as the functions of the disk masses and BH spin parameters to contrast the observational jet luminosities of GRBs. The results show that, although the neutrino annihilation processes could account for most of GRBs, the BZ mechanism is more effective, especially for long-duration GRBs. Actually, if the energy of afterglows and flares of GRBs is included, the distinction between these two mechanisms is more significant. Furthermore, massive disk mass and high BH spin are beneficial to power high luminosities of GRBs. Finally, we discuss possible physical mechanisms to enhance the disk mass or the neutrino emission rate of NDAFs and relevant difference between these two mechanisms.

\end{abstract}


\section{Introduction}

The progenitors of short-duration and long-duration gamma-ray bursts (SGRBs and LGRBs) are respectively believed to result from the merger of two compact objects, i.e., two neutron stars (NSs) or a black hole (BH) and a NS, and the core collapse of a massive star \citep[see, e.g.,][]{Eichler1989,Paczynski1991,Narayan1992,Woosley1993,Paczynski1998}. A BH hyperaccretion system is expected to form in the center of gamma-ray bursts (GRBs). This geometrically thick and extremely optically thick hyperaccretion disk with high density and temperature is named the neutrino-dominated accretion flow (NDAF), which has been widely studied including the researches on the time-independent radial structure and relevant Blandford-Zanjek \citep[BZ,][]{Blandford1977} or neutrino luminosity \citep[e.g.,][]{Popham1999,DiMatteo2002,Kohri2002,Kohri2005,Gu2006,Liu2007,Kawanaka2007,Lei2009,Kawanaka2013,Li2013,Luo2013,Xue2013}, on the time-independent vertical structure and relevant neutrino luminosity \citep[e.g.,][]{Liu2008,Liu2010a,Liu2012a,Liu2013,Liu2014,Liu2015a}, on the applications to GRBs observations \citep[e.g.,][]{Reynoso2006,Lazzati2008,Liu2010b,Liu2012b,Barkov2011,Sun2012,Hou2014a,Hou2014b,Liu2015b}, and on the various time-dependent simulations \citep[e.g.,][]{Ruffert1999,Lee2004,Lee2009,Janiuk2013}.

The magnetic field plays an important role in astrophysics, especially in high-energy astrophysics. Without exception, it is also a key role in GRBs. Two scenarios are often discussed on the central engine of GRBs. Firstly, as mentioned above, the hyperaccretion system should launch a relativistic jet to power a GRB, which origin mechanisms include neutrino-antineutrino annihilation \citep[e.g.,][]{Popham1999,Liu2007,Liu2010a} and magnetohydrodynamical process such as BZ process \citep[e.g.,][]{Lee2000a,Lee2000b,DiMatteo2002,Kawanaka2013}. Secondly, apart from NDAF models, the events of the two NSs mergers or core collapses may produce a massive proto-magnetars to power GRBs and their X-ray flares \citep[e.g.,][]{Dai2006,Metzger2011,Gao2013,Kumar2015,Lai2015,Wang2015}. Although the present GRBs observations cannot clearly tell us which candidate certainly exists in the center of GRBs, yet for the BH hyperaccretion process, it is still possible to contrast and identify the BZ mechanism and neutrino annihilation by the actual measurements of GRBs.

In this paper, we focus on the comparison of the jet luminosity driven by the BZ mechanism and neutrino annihilation by means of the observational data of SGRBs and LGRBs. In Section 2, we present our NDAF model and give the analytic formulae of the BZ jet power and neutrino pair annihilation. We apply these two mechanisms to explain the observation data of GRBs in order to discuss the feasibilities in Section 3. Conclusions and discussion are in Section 4.

\section{Models}

In \citet{Xue2013}, we investigated one-dimensional global solutions of NDAFs, taking account of general relativity in Kerr metric, neutrino physics and nucleosynthesis more precisely than most previous works \citep[e.g,][]{Kohri2002,Kohri2005,Liu2007,Kawanaka2007}. In details, we considered that the total optical depth for neutrinos including scattering of electrons and nucleons and absorption through four terms, i.e., the Urca processes, electron-positron pair annihilation, nucleon-nucleon bremsstrahlung, and plasmon decay \citep[e.g.,][]{DiMatteo2002,Liu2007}. In order to allow for a transition from the optically thin to optically thick regions, a bridging formula of free protons and neutrons was established by the relations of the reaction rates in the $\beta$ processes. We applied the proton-rich material in a state of nuclear statistical equilibrium \citep{Seitenzahl2008} to NDAF model, which is suitable for almost all the range of the electron fraction. The complicated and detailed balance are included under the equilibrium of the chemical potential.

We calculated sixteen solutions with different characterized mass accretion rates and BH spins, and exhibited the radial distributions of various physical properties in NDAFs. The results showed that the gas pressure and the neutrino cooling always become dominant in the inner region for large accretion rates, and the electron degeneracy should not be ignored. Electron fraction is always about 0.46 in the outer region, and the inner, middle, and outer regions are always dominated by the free nucleons, $\rm ^4He$, and $\rm ^{56}Fe$.

We also calculated the neutrino luminosity and annihilation luminosity by considering the influence of neutrino trapping and proportion of heavy nuclei. Even in this case, most of the solutions show the adequate annihilation luminosities to satisfy the requirement of the mean luminosity of GRBs. Therefore, we would like to estimate the jet luminosity through the candidate BZ process or  neutrino annihilation basing our results on \citet{Xue2013}.

\subsection{BZ Luminosity and Neutrino Annihilation Luminosity}

\citet{Blandford1977} presented that the rotational energy of a BH can be tremendously extracted to power a Poynting jet via a large-scale poloidal magnetic field threading the horizon of the BH. The BZ luminosity can be estimated as \citep[e.g.,][]{Krolik2011,Kawanaka2013}
\beq L_{\rm BZ} = f(a_*)c R_{\rm g}^2 \frac{B_{\rm in}^2}{8 \pi}, \eeq
where $a_*$ is the dimensionless BH spin, $f (a_*)$ is a factor depending on the specific configuration of the magnetic field \citep[e.g.,][]{Blandford1977,Tchekhovskoy2008,Kawanaka2013}, $R_{\rm g} =2 GM/c^2$ is the Schwarzschild radius, $M$ is the mass of the BH, and $B_{\rm in}$ is the poloidal magnetic field strength near the horizon. Moreover, the informations of the magnetic field configuration are included in $f (a_*)$. For the specified magnetic field geometries, the analytical $f (a_*)$ were attempted \citep[e.g.,][]{Blandford1977,Tchekhovskoy2008}, but these configurations were not required to be consistent with the dynamics of the accretion disk. Many two- or three-dimensional MHD simulations have investigated how a large-scale vertical magnetic field evolves with an accretion disk and which configurations can power a jet \citep[e.g.,][]{McKinney2004,Beckwith2008,Beckwith2009,McKinney2009}, yet the form of $f (a_*)$ is still unclear. We only know that it is an increasing function of $a_*$, and its range is from a small number to $\sim 1$ \citep{Hawley2006}. We consider that the structure of the magnetic field is positively important for the BZ luminosity and the structure of the disk. As simplification, we assume $f (a_*)=1$ as well as that in \citet{Kawanaka2013}, which is suitable for the fast-spinning BH in the center of GRBs.

The magnetic field energy can be estimated by the disk pressure near the horizon $P_{\rm in}$ presented as
\beq \beta_h \frac{B_{\rm in}^2}{8\pi}= P_{\rm in}, \eeq
where $\beta_h$ is the ratio of the midplane pressure near the horizon of the BH to the magnetic pressure in the stretched horizon. Following \citet{Kawanaka2013}, we also adopt $\beta_h$ to unity.

Following \citet{Xue2013}, we set the BH mass $M=3~M_\odot$ and the constant viscosity parameter $\alpha=0.1$, which are the typical settings for GRBs. The analytic formula of the disk pressure near the horizon is a function of BH spin $a_*$ ($0\leq a_*<1$) and dimensionless mass accretion rate $\dot{m}$ ($\dot{m}\equiv \dot{M}/M_\odot~\rm s^{-1}$, and $\dot{M}$ is the accretion rate), which can be approximated from the data of \citet{Xue2013} as
\beq \log P_{\rm in}~(\rm{erg\ cm^{-3}})&\approx&30.0+1.22a_*+1.00\log\dot{m}. \eeq

Under the same conditions stated above and considering the effect of the neutrino trapping, the analytic formula of the neutrino annihilation luminosity above the accretion flow can be written as a function of BH spin and accretion rate \citep{Xue2013}
\beq \log L_{\nu \bar{\nu}}~(\rm{erg\ s^{-1}})&\approx& 49.5+2.45a_*+2.17\log\dot{m}. \eeq
We have verified that these analytic formulae are almost applicable for all the mass accretion rate higher than the ignition accretion rate.

Additionally, we noticed that the BH mass and viscosity parameter have some significant effects on the structure and components of the disk \citep[e.g.,][]{Popham1999,Chen2007}. It is noteworthy that the variation of the viscosity parameter has little effects on neutrino emission rate in the innermost region ($\lesssim 10~R_{\rm g}$) of the disk. In this region, the free protons and neutrons are dominant, so the neutrino reactions related to the neutrino emission mainly occur here \citep[e.g.,][]{Popham1999,Liu2007,Li2013,Xue2013}. In these cases with different viscosity parameters, the numbers of the launched neutrinos are roughly equal because of the similar temperatures in the innermost regions, because the cooling rate of Urca and other processes are mainly related to the temperature of the disk \citep[e.g.,][]{DiMatteo2002}. Thus the neutrino luminosity is almost independent of the viscosity value, much less the annihilation luminosity. \citet{Zalamea2011} also claimed that the uncertainty in viscosity parameter has almost no effect on annihilation luminosity. Furthermore, we calculated how the annihilation luminosity was determined by the fundamental parameters of the BH accretion system \citep{Wang2009}. It is shown that the annihilation luminosity is almost independent of $\alpha$, and is not significantly related to the BH mass if it is set as several solar mass. Here only the BH spin and accretion rate are taken into account.

\subsection{Methods}

For SGRBs and LGRBs, if we know the isotropic luminosity $L_{\rm iso}$ and jet opening angle $\theta_{\rm jet}$ or bulk Lorentz factor $\Gamma$, the jet luminosity $L_{\rm jet}$ can be expressed as
\beq L_{\rm jet}=L_{\rm iso}(1- \cos \theta_{\rm jet}) \approx L_{\rm iso} \Gamma^{-2}. \eeq
Moreover, the average accretion rate $\dot{M}$ can be estimated by
\beq \dot{M}\approx M_{\rm disk}(1+z)/T_{90}, \eeq
where $M_{\rm disk}$, $T_{90}$, and $z$ are the mass of the disk, the duration and redshift of GRBs, respectively. Hereafter, we would like to use the dimensionless mass of the disk defined as $m_{\rm disk}\equiv M_{\rm disk}/M_\odot$. Once the disk mass and BH spin are given, we can estimate the BZ jet luminosity and neutrino annihilation luminosity by Equations (1) and (4). For convenience, we define two dimensionless parameter $\tau_1$ and $\tau_2$ as
\beq \tau_1 = \log (L_{\rm jet}/L_{\nu \bar{\nu}}),\\
\tau_2 = \log (L_{\rm jet}/L_{\rm BZ}), \eeq
to illustrate which mechanism is suitable for explaining SGRBs or LGRBs. It is difficult to estimate the jet opening angle independent of models, so we have to replace the angles by the calculative Lorentz factors related to the fireball model. We consider that the jet luminosity calculated from Equation (5) can also generally evaluate the BZ power.

\citet{Eichler1989} proposed that the mergers of two NSs might be the candidates for powering SGRBs. Subsequently, \citet{Ruffert1998} reported results of three-dimensional Newtonian hydrodynamical simulations of the collision of two identical neutron stars with  mass $\sim 1.6~M_\odot$. One of endings might be a BH $\sim 2.5 M_\odot$ surrounded by a disk $\sim 0.1-0.2~M_\odot$. The merger of a NS and a stellar-mass BH also can produce SGRBs \citep{Paczynski1991,Narayan1992}. The simulations showed that the larger mass of the disk was formed, $\sim 0.5 M_\odot$ \citep[e.g.,][]{Kluzniak1998,Lee1999,Popham1999,Liu2012b}. \citet{Woosley1993} suggested that the core collapsar could power LGRBs, and the stellar-mass BH hyperaccretion systems might arise in the center \citep[e.g.,][]{MacFadyen1999,Zhang2003}, whose disk mass was about several solar mass \citep[e.g.,][]{Popham1999}. In summery, we take the typical dimensionless mass of the disk as 0.2, 0.5 and 1, 3 for SGRBs and LGRBs, respectively.

The BH spin parameter is an important ingredient for the occurrence of GRBs \citep[e.g.,][]{Janiuk2008,Liu2010a,Xue2013}, so we choose the large spin parameters as 0.5 and 0.9 in the investigated cases, which are consistent with $f(a_*)=1$. The detailed definition of $\tau$ for two classes of GRBs, and different typical disk masses and BH spin parameters are shown in Table 1.

\section{Results}

The parameters of 21 SGRBs and 55 LGRBs in our sample are presented in Tables 2 and 3, respectively, which include the redshift $z$, GRB duration $T_{90}$, isotropic mean gamma-ray luminosity $L_{\rm iso}$, initial Lorentz factor $\Gamma$, and dimensionless parameters $\tau$ with different disk masses and BH spin parameters as shown in Table 1. The data are employed from \citet{Fan2011}, \citet{Lv2012}, \citet{Berger2014} and \citet{Tang2014}, and the Lorentz factor of SGRBs cited from \citet{Berger2014} are estimated by the peak time of the afterglow $t_{\rm peak}$ as well as methods in \citet{Fan2011} or \citet{Lv2012}. Although the calculative Lorentz factors are related to the fireball model, we consider that the jet luminosity can also roughly evaluate the BZ power.

In these tables, the local durations of SGRBs are shorter than 2 s, conversely, durations of LGRBs are much longer than 2 s. The Lorentz factors and isotropic luminosities of LGRBs are generally larger than those of SGRBs. The parameters $\tau$ are calculated by Equations (1-8) to measure which mechanism is more effective and which set of parameters is more suitable for a certain GRB than the others as well as Figures 1 and 2.

Figures 1 and 2 display the distributions of $\tau$ for SGRBs and LGRBs under the BZ mechanism or neutrino annihilation with the different BH spins and disk masses, respectively. No matter what class or what mechanism the GRB is, the larger BH spin or larger disk mass can more effectively produce the jet luminosity of GRBs. For example, for the data of LGRBs and BZ mechanism, there are only 5 and 3 GRBs with $\tau > -3$ for the case of $m_{\rm disk} =1$ and $a_*=0.5$ and the case of $m_{\rm disk} =3$ and $a_*=0.5$ as the black solid lines shown in Figures 2 (a) and (b). That means the larger disk masses are more conducive to satisfy the energy requirements of GRBs. Also for the data of LGRBs and BZ mechanism, there are only 3 and 1 GRBs with $\tau > -3$ for the case of $m_{\rm disk} =3$ and $a_*=0.5$ and the case of $m_{\rm disk} =3$ and $a_*=0.9$ as the black solid line and dashed line shown in Figure 2 (b). That means the larger BH spin are more conducive to power GRBs. From the number distributions in the histograms, the effect of BH spin is more significant than that of disk mass for SGRBs and LGRBs.

More importantly, we should focus on the comparison between the black solid (or dashed) lines and red solid (or dashed) lines in these figures, i.e., comparison between BZ mechanisms and neutrino annihilation. Obviously, despite of the high BH spin and disk mass benefiting the high annihilation luminosity, overall the BZ power is more effective than the annihilation process for both SGRBs and LGRBs. For the SGRB cases of $m_{\rm disk} =0.2$ and $a_*=0.5$, there are 3 and 0 GRBs with $\tau > 0$ respectively correspond the annihilation processes and BZ mechanism, which means that the annihilation process absolutely cannot be the candidate of the central engine in the these three SGRBs. For the SGRB cases of $m_{\rm disk} =0.5$ and $a_*=0.9$, the mean $\tau$ of the BZ mechanism is also much lower than that of the annihilation process. More seriously, For LGRB cases, there are more instances cannot be explained by annihilation process.

If the neutrino annihilation process is thought to power GRBs, the jet mean power outputting is a fraction of the neutrino annihilation luminosity, and the scaling factor is considered as about 0.3 \citep[e.g.,][]{Aloy2005,Fan2011,Liu2015b}. Actually, the more serious problem is that the energies of the afterglows and flares of GRBs are not yet included in our calculations. It is doing so in order to avoid the more dependence of model and the uncertainty of the observational data. As we know, the energy of afterglow may be equal to or much more than that of the prompt emission, so the scaling factor may be counteracted. We ignore the efficiency problem in the radiation process that the annihilation luminosity is transferred into the jet luminosity. Thus, even $\tau<0$ may not go far enough if the energy of afterglows and flares are about one order of magnitude higher than that of the prompt emission. As shown in figures, the distinction between two mechanisms is more significant and neutrino annihilation is more ineffective, if we consider $\tau<-1$ is reasonable.

Furthermore, the jet luminosities of GRBs 060313, 090323 and 091024 are quite higher than the other samples, so even for the BZ mechanism, higher BH spin or disk mass is required in these cases. In conclusion, the BZ mechanism is more powerful than the annihilation process. We should not only investigate which mechanism is more effective, but also find more evidences to distinguish which mechanism really exists in the center of GRBs.

\section{Conclusions and Discussion}

We use the analytic solutions from \citet{Xue2013} and ignore the effects of the magnetic field configuration to estimate the BZ and neutrino annihilation luminosities as the functions of the disk masses and BH spin parameters, then to contrast the observational jet luminosity of GRBs. Our results show that, although the neutrino annihilation processes could account for most of GRBs, the BZ mechanism is more effective, especially for LGRBs. Furthermore, massive disk mass and high BH spin are beneficial to power high luminosities of GRBs. Actually, there are some physical possibilities to enhance neutrino annihilation luminosity, such as increasing the disk mass or the neutrino emission rate of NDAF, and two mechanisms have possible different observational effects, which we would like to discuss below.

\subsection{Disk Mass for SGRBs}

The massive NSs are widely discussed in observations and theoretical models \citep[e.g.,][]{Morrison2004,Dai2006,Li2012,Strader2015}. In the past several years, the massive NSs, i.e., $\sim 2~M_\odot$ have been discovered by the dynamic measurements \citep{Demorest2010,Antoniadis2013}. Although their companions are white dwarfs, we cannot completely rule out the possibilities that the massive NSs exist in the binaries constituted by  two NSs or a BH and a NS. Thus a massive disk $\geq 0.5~M_\odot$ for SGRBs may form a coalescence of a BH (or a NS) with a massive NS, i.e., $\geq 2~M_\odot$, and the high neutrino annihilation luminosity could be expected. Recently, \citet{Strader2015} reported that through multiwavelength observations, a \emph{Fermi}-LAT unidentified gamma-ray source 1FGL J1417.7-4407 may contain a massive NS (nearly 2 $M_\odot$) and a $\sim 0.35~M_\odot$ giant secondary with a 5.4 day period.

Moreover, since the mergers of compact objects may produce LGRBs due to the radial transfer of the angular momentum of the massive disk \citep{Liu2012b} and off-axis jet from collapsars may trigger SGRBs \citep{Lazzati2010}, the shortage of the accretion matter does not exhibit if some high-luminous SGRBs originate from collapsars.

Of course, we cannot neglect that the outflow from NDAFs influences the disk mass and neutrino luminosity. \citet{Janiuk2013} studied two-dimensional relativistic NDAF models and resulted that the neutrino-cooled torus launched a powerful mass outflow, which contributed to the total neutrino luminosity and mass loss from the system. We investigated the radial outflow for the angular momentum transfer, which also caused the mass loss from the disk \citep{Liu2012b}. Furthermore, for NDAFs, the high accretion rate leads to the violent evolution of the mass and spin of the central BH, which should also affect the neutrino radiation rate from the disk. In future, we will calculate the co-evolution of a BH and its surrounding NDAF to depict the more authentic pictures that indeed what happens in the center of GRBs.

\subsection{Neutrino Emission Rate}

Recently, \citet{Jiang2014} studied BH super-Eddington accretion flows using a global three-dimensional radiation magneto-hydrodynamical simulation. They found that the vertical advection of radiation caused by magnetic buoyancy can much effectively transport energy to increase photons emission. The similar mechanism may exist in NDAF models, which can increase neutrino emission rate. Similarly, \citet{Liu2015a} investigated the effects of the vertical convection on the structure and luminosity of the NDAF. Since the gas and neutrinos are carried to the nearby disk surface, and the convective energy transferred in the vertical direction can be effective to suppress the advection, the neutrino luminosity and annihilation luminosity are increased more than an order of magnitude for $\dot{M}\geq 1~M_\odot~\rm s^{-1}$, which is conducive to achieve the energy requirement of GRBs.

Furthermore, \citet{Lei2009} and \citet{Luo2013} studied NDAF model with the magnetic coupling between the inner disk and BH. In this framework, the angular momentum and the energy can be transferred from the horizon of the BH to the disk. Thus the neutrino luminosity and relevant annihilation luminosity can be significantly enhanced to power high luminosities of GRBs.

\subsection{Polarization}

If the jet powering the prompt emission and late X-ray flares of GRBs is launched by magnetic fields, GRBs are expected to be the astronomical candidate sources of linearly polarized \citep[e.g.,][]{Fan2005}. \citet{Mundell2013} reported the detection of degrees of linear polarization about 28 percent in the afterglow of LGRB GRB 120308, which might indicate that large-scale magnetic fields were presented in the GRB jets \citep{Lai2015}. Whether high polarizations do exist in all GRB jets, future GRBs observations by the POLAR detector may further test the possibility and identify the BZ mechanism and neutrino annihilation process.

\acknowledgments

We thank the anonymous referee for very useful suggestions and comments. This work was supported by the National Basic Research Program of China (973 Program) under grant 2014CB845800, the National Natural Science Foundation of China under grants 11222328, 11233006, 11333004, 11373002, 11473022, and U1331101, the CAS Open Research Program of Key Laboratory for the Structure and Evolution of Celestial Objects under grants OP201305 and OP201403.

\clearpage

\begin{table*}\label{table1}
\caption{Definition of $\tau$}
\centering
\begin{tabular}{lllllllllllll}
\hline \hline
 &$\tau_{11}^S$ & $\tau_{12}^S$& $\tau_{21}^S$ & $\tau_{22}^S$ & $\tau_{13}^S$ & $\tau_{14}^S$ & $\tau_{23}^S$ & $\tau_{24}^S$\\
\hline
$\rm Class^1$           &	S 	&	S 	&	S 	&	S 	&	S 	&	S 	&	S 	&	S	\\
$\rm Mechanism^2$       &	1 	&	1 	&	2 	&	2 	&	1 	&	1 	&	2 	&	2 	\\
$a_*$	        &	0.5 	&	0.5 	&	0.5 	&	0.5 	&	0.9 	&	0.9 	&	0.9 	&	0.9 	\\
$m_{\rm disk}$	&	0.2 	&	0.5 	&	0.2 	&	0.5 	&	0.2 	&	0.5 	&	0.2 	&	0.5 	\\
\hline
&$\tau_{11}^L$ & $\tau_{12}^L$ & $\tau_{21}^L$ & $\tau_{22}^L$ & $\tau_{13}^L$ & $\tau_{14}^L$ & $\tau_{23}^L$ & $\tau_{24}^L$\\
\hline
Class           &	L 	&	L 	&	L 	&	L 	&	L 	&	L 	&	L 	&	L 	\\
Mechanism       &	1 	&	1 	&	2 	&	2 	&	1 	&	1 	&	2 	&	2 	\\
$a_*$	        &	0.5 	&	0.5 	&	0.5 	&	0.5 	&	0.9 	&	0.9 	&	0.9 	&	0.9 	\\
$m_{\rm disk}$	&	1 	&	3 	&	1 	&	3 	&	1 	&	3 	&	1 	&	3 	\\
\hline
\tablenotetext{1}{Class: S - SGRBs; L - LGRBs.}
\tablenotetext{2}{Mechanism: 1 - neutrino annihilation process; 2 - BZ mechanism.}
\end{tabular}
\end{table*}

\clearpage

\begin{deluxetable}{lllllcllllllll}\label{table2}
\centering
\tabletypesize{\scriptsize}
\tablewidth{0pt}
\tablecaption{Data of SGRBs}
\tablehead{\colhead{$\rm SGRB$} & \colhead{$z$} & \colhead{$T_{90}(\rm s)$} & \colhead{$L_{\rm iso}(10^{52}~\rm erg~s^{-1})$} & \colhead{$\Gamma$} & \colhead{Ref.$^1$} & \colhead{$\tau_{11}^S$} & \colhead{$\tau_{12}^S$} & \colhead{$\tau_{21}^S$} & \colhead{$\tau_{22}^S$} & \colhead{$\tau_{13}^S$} & \colhead{$\tau_{14}^S$} & \colhead{$\tau_{23}^S$} & \colhead{$\tau_{24}^S$}}
\startdata
021211	&	1.006 	&	2.3 	&	0.98 	&	195 	& a	& -1.67 	&	-2.53 	&	-4.24 	&	-4.64 	&	-2.65 	&	-3.51 	&	-4.73 	 &	-5.13 	\\
040924	&	0.858 	&	2.39 	&	1.17 	&	490 	& a	& -2.28 	&	-3.15 	&	-4.92 	&	-5.31 	&	-3.26 	&	-4.13 	&	-5.40 	 &	-5.80 	\\
050709	&	0.16 	&	0.07 	&	0.11 	&	4.76 	& b	& -2.15 	&	-3.02 	&	-3.23 	&	-3.63 	&	-3.13 	&	-4.00 	&	-3.71 	 &	-4.11 	\\
050724	&	0.257 	&	3 	    &	0.02 	&	5 	    & b	& 0.44 	    &	-0.43 	&	-2.51 	&	-2.91 	&	-0.54 	&	-1.41 	&	-3.00 	 &	-3.39 	\\
051210	&	1.3 	&	1.3	    &	0.63 	&	39.34 	& c	& -1.14 	&	-2.00 	&	-3.35 	&	-3.75 	&	-2.12 	&	-2.98 	&	-3.84 	 &	-4.24 	\\
051221A	&	0.55 	&	1.4 	&	0.27 	&	10 	    & b	& 0.12 	    &	-0.74 	&	-2.33 	&	-2.73 	&	-0.86 	&	-1.72 	&	-2.82 	 &	-3.22 	\\
060313	&	1.7 	&	0.7 	&	31.09 	&	7.05 	& c	& 1.32 	    &	0.45 	&	-0.50 	&	-0.90 	&	0.34 	&	-0.53 	&	-0.99 	 &	-1.39 	\\
061006	&	0.4377 	&	0.4	    &	0.75 	&	9.09 	& b	& -0.45 	&	-1.32 	&	-2.31 	&	-2.70 	&	-1.43 	&	-2.30 	&	-2.79 	 &	-3.19 	\\
061201	&	0.11 	&	0.8 	&	0.01 	&	3.52 	& c	& -0.50 	&	-1.37 	&	-2.84 	&	-3.24 	&	-1.48 	&	-2.35 	&	-3.33 	 &	-3.73 	\\
070714B	&	0.92 	&	3 	    &	0.08 	&	12.5 	& b	& -0.10 	&	-0.96 	&	-2.83 	&	-3.23 	&	-1.08 	&	-1.94 	&	-3.31 	 &	-3.71 	\\
070724A	&	0.46 	&	0.4 	&	0.06 	&	3.5 	& c	& -0.76 	&	-1.62 	&	-2.60 	&	-3.00 	&	-1.74 	&	-2.60 	&	-3.09 	 &	-3.49 	\\
070809	&	0.47 	&	1.3 	&	0.06 	&	8.04 	& c	& -0.33 	&	-1.20 	&	-2.78 	&	-3.17 	&	-1.31 	&	-2.18 	&	-3.26 	 &	-3.66 	\\
071112c	&	0.82 	&	0.3 	&	0.51 	&	6	    & c	& -0.75 	&	-1.62 	&	-2.34 	&	-2.74 	&	-1.73 	&	-2.60 	&	-2.83 	 &	-3.22 	\\
071227	&	0.381 	&	1.8 	&	0.04 	&	11.11 	& b	& -0.40 	&	-1.27 	&	-3.04 	&	-3.44 	&	-1.38 	&	-2.25 	&	-3.53 	 &	-3.92 	\\
080905A	&	0.12 	&	1 	    &	0.01 	&	11.18 	& c	& -1.69 	&	-2.55 	&	-4.13 	&	-4.53 	&	-2.67 	&	-3.53 	&	-4.62 	 &	-5.02 	\\
090426	&	2.609 	&	1.25 	&	0.87 	&	16.67 	& b	& -0.71 	&	-1.58 	&	-2.68 	&	-3.08 	&	-1.69 	&	-2.56 	&	-3.17 	 &	-3.56 	\\
090510	&	0.903 	&	0.3 	&	40.3 	&	123 	& b	& -1.52 	&	-2.39 	&	-3.09 	&	-3.49 	&	-2.50 	&	-3.37 	&	-3.57 	 &	-3.97 	\\
090515	&	0.4 	&	0.04 	&	0.30 	&	30.99 	& c	& -4.07 	&	-4.94 	&	-4.77 	&	-5.16 	&	-5.05 	&	-5.92 	&	-5.25 	 &	-5.65 	\\
100117A	&	0.92 	&	0.3 	&	1.30 	&	37.23 	& c	& -1.98 	&	-2.85 	&	-3.54 	&	-3.94 	&	-2.96 	&	-3.83 	&	-4.03 	 &	-4.43 	\\
100625A	&	0.45 	&	0.3 	&	0.57 	&	19.07 	& c	& -1.50 	&	-2.36 	&	-3.20 	&	-3.60 	&	-2.48 	&	-3.34 	&	-3.69 	 &	-4.08 	\\
100816A	&	0.8035 	&	2.8 	&	0.37 	&	100 	& b	& -1.22 	&	-2.08 	&	-3.95 	&	-4.35 	&	-2.20 	&	-3.06 	&	-4.44 	 &	-4.83 	\\
\enddata
\tablenotetext{1}{References: (a) L{\"u} et al. (2012); (b) Fan \& Wei (2011); (c) Berger (2014).}
\end{deluxetable}

\clearpage

\begin{deluxetable}{lllllcllllllll}\label{table3}
\centering
\tabletypesize{\scriptsize}
\tablewidth{0pt}
\tablecaption{Data of LGRBs}
\tablehead{\colhead{$\rm LGRB$} & \colhead{$z$} & \colhead{$T_{90}(\rm s)$} & \colhead{$L_{\rm iso}(10^{52}~\rm erg~s^{-1})$} & \colhead{$\Gamma$} & \colhead{Ref.$^1$} &\colhead{$\tau_{11}^L$} & \colhead{$\tau_{12}^L$} & \colhead{$\tau_{21}^L$} & \colhead{$\tau_{22}^L$} & \colhead{$\tau_{13}^L$} & \colhead{$\tau_{14}^L$} & \colhead{$\tau_{23}^L$} & \colhead{$\tau_{24}^L$}}
\startdata
990123	&	1.61 	&	63.3 	&	9.44 	&	600 	& a	& -0.30 	&	-1.34 	&	-3.61 	&	-4.09 	&	-1.28 	&	-2.32 	&	-4.10 	 &	-4.57 	\\
050401	&	2.9 	&	33 	    &	4.14 	&	590 	& a	& -1.64 	&	-2.67 	&	-4.41 	&	-4.89 	&	-2.62 	&	-3.65 	&	-4.90 	 &	-5.37 	\\
050525A	&	0.606 	&	8.8 	&	1.74 	&	229 	& a	& -1.60 	&	-2.64 	&	-4.15 	&	-4.63 	&	-2.58 	&	-3.62 	&	-4.64 	 &	-5.12 	\\
050730	&	3.97 	&	155 	&	0.29 	&	201 	& a	& -0.63 	&	-1.66 	&	-4.06 	&	-4.54 	&	-1.61 	&	-2.64 	&	-4.55 	 &	-5.03 	\\
050801	&	1.56 	&	20 	    &	0.12 	&	420 	& a	& -2.95 	&	-3.99 	&	-5.69 	&	-6.16 	&	-3.93 	&	-4.97 	&	-6.17 	 &	-6.65 	\\
050820A	&	2.615 	&	600 	&	0.58 	&	282 	& a	& 0.96 	    &	-0.08 	&	-3.33 	&	-3.81 	&	-0.02 	&	-1.06 	&	-3.82 	 &	-4.29 	\\
050922C	&	2.198 	&	4.54 	&	3.56 	&	274 	& a	& -2.72 	&	-3.75 	&	-4.58 	&	-5.06 	&	-3.70 	&	-4.73 	&	-5.07 	 &	-5.55 	\\
060210	&	3.91 	&	220 	&	0.93 	&	264 	& a	& -0.02 	&	-1.05 	&	-3.64 	&	-4.11 	&	-1.00 	&	-2.03 	&	-4.12 	 &	-4.60 	\\
060418	&	1.49 	&	52 	    &	0.48 	&	263 	& a	& -1.02 	&	-2.06 	&	-4.25 	&	-4.73 	&	-2.00 	&	-3.04 	&	-4.74 	 &	-5.21 	\\
060605	&	3.8 	&	19 	    &	0.63 	&	197 	& a	& -2.22 	&	-3.25 	&	-4.60 	&	-5.08 	&	-3.20 	&	-4.23 	&	-5.09 	 &	-5.57 	\\
060607A	&	3.082 	&	100 	&	0.37 	&	296 	& a	& -1.08 	&	-2.12 	&	-4.40 	&	-4.87 	&	-2.06 	&	-3.10 	&	-4.88 	 &	-5.36 	\\
060904B	&	0.703 	&	192 	&	0.01 	&	108 	& a	& -0.53 	&	-1.57 	&	-4.62 	&	-5.10 	&	-1.51 	&	-2.55 	&	-5.11 	 &	-5.58 	\\
060908	&	2.43 	&	19.3 	&	1.90 	&	304 	& a	& -1.78 	&	-2.82 	&	-4.35 	&	-4.83 	&	-2.76 	&	-3.80 	&	-4.83 	 &	-5.31 	\\
061007	&	1.262 	&	75 	    &	3.16 	&	436 	& a	& -0.20 	&	-1.24 	&	-3.67 	&	-4.15 	&	-1.18 	&	-2.22 	&	-4.16 	 &	-4.63 	\\
061121	&	1.314 	&	81 	    &	0.75 	&	175 	& a	& 0.01 	    &	-1.02 	&	-3.48 	&	-3.96 	&	-0.97 	&	-2.00 	&	-3.97 	 &	-4.44 	\\
070110	&	2.352 	&	89 	    &	0.21 	&	127 	& a	& -0.52 	&	-1.56 	&	-3.87 	&	-4.35 	&	-1.50 	&	-2.54 	&	-4.36 	 &	-4.84 	\\
070318	&	0.84 	&	63 	    &	0.04 	&	143 	& a	& -1.08 	&	-2.12 	&	-4.57 	&	-5.04 	&	-2.06 	&	-3.10 	&	-5.05 	 &	-5.53 	\\
070411	&	2.954 	&	101	    &	0.39 	&	208 	& a	& -0.72 	&	-1.75 	&	-4.05 	&	-4.53 	&	-1.70 	&	-2.73 	&	-4.54 	 &	-5.01 	\\
070419A	&	0.97 	&	112 	&	0.0042 	&	91   	& a	& -1.21 	&	-2.25 	&	-4.95 	&	-5.43 	&	-2.19 	&	-3.23 	&	-5.44 	 &	-5.92 	\\
071003	&	1.1 	&	148 	&	0.97 	&	283 	& a	& 0.37 	    &	-0.67 	&	-3.48 	&	-3.96 	&	-0.61 	&	-1.65 	&	-3.97 	 &	-4.44 	\\
071010A	&	0.98 	&	6 	    &	0.04 	&	101 	& a	& -3.06 	&	-4.09 	&	-5.31 	&	-5.78 	&	-4.04 	&	-5.07 	&	-5.79 	 &	-6.27 	\\
071010B	&	0.947 	&	35.74 	&	0.14 	&	209 	& a	& -1.48 	&	-2.51 	&	-4.64 	&	-5.12 	&	-2.46 	&	-3.49 	&	-5.13 	 &	-5.61 	\\
071031	&	2.692 	&	150.49 	&	0.10 	&	133 	& a	& -0.50 	&	-1.53 	&	-4.07 	&	-4.54 	&	-1.48 	&	-2.51 	&	-4.55 	 &	-5.03 	\\
080129	&	4.394 	&	48 	    &	0.79 	&	65  	& a	& -0.39 	&	-1.43 	&	-3.19 	&	-3.67 	&	-1.37 	&	-2.41 	&	-3.68 	 &	-4.15 	\\
080319B	&	0.937 	&	57 	    &	4.49 	&	580 	& a	& -0.41 	&	-1.45 	&	-3.82 	&	-4.30 	&	-1.39 	&	-2.43 	&	-4.30 	 &	-4.78 	\\
080319C	&	1.95 	&	29.55 	&	2.25 	&	228 	& a	& -0.92 	&	-1.95 	&	-3.78 	&	-4.25 	&	-1.90 	&	-2.93 	&	-4.26 	 &	-4.74 	\\
080330	&	1.51 	&	61 	    &	0.02 	&	104 	& a	& -1.52 	&	-2.56 	&	-4.83 	&	-5.31 	&	-2.50 	&	-3.54 	&	-5.32 	 &	-5.79 	\\
080413B	&	1.1 	&	8.0	    &	0.47 	&	128 	& a	& -2.01 	&	-3.04 	&	-4.37 	&	-4.85 	&	-2.99 	&	-4.02 	&	-4.86 	 &	-5.34 	\\
080603A	&	1.688 	&	150 	&	0.03 	&	88  	& a	& -0.39 	&	-1.43 	&	-4.12 	&	-4.60 	&	-1.37 	&	-2.41 	&	-4.61 	 &	-5.09 	\\
080710	&	0.845 	&	120 	&	0.01 	&	63  	& a	& -0.31 	&	-1.35 	&	-4.12 	&	-4.60 	&	-1.29 	&	-2.33 	&	-4.61 	 &	-5.08 	\\
080810	&	3.35 	&	108 	&	1.21 	&	409 	& a	& -0.84 	&	-1.87 	&	-4.16 	&	-4.64 	&	-1.82 	&	-2.85 	&	-4.64 	 &	-5.12 	\\
080916C	&	4.35 	&	66 	    &	71.33 	&	1130    & a	& -0.61 	&	-1.65 	&	-3.57 	&	-4.05 	&	-1.59 	&	-2.63 	&	 -4.06 	&	-4.54 	\\
081203A	&	2.1 	&	223 	&	0.24 	&	219 	& a	& 0.004 	    &	-1.03 	&	-3.86 	&	-4.33 	&	-0.98 	&	-2.01 	&	-4.34 	 &	-4.82 	\\
090313	&	3.375 	&	78 	    &	0.18 	&	136  	& a	& -1.02 	&	-2.06 	&	-4.17 	&	-4.65 	&	-2.00 	&	-3.04 	&	-4.66 	 &	-5.14 	\\
090323	&	3.568 	&	150 	&	12.15 	&	110 	& a	& 1.57 	    &	0.53 	&	-1.89 	&	-2.37 	&	0.59 	&	-0.45 	&	-2.38 	 &	-2.86 	\\
090328A	&	0.736 	&	70 	    &	0.57 	&	540 	& a	& -0.95 	&	-1.99 	&	-4.52 	&	-4.99 	&	-1.93 	&	-2.97 	&	-5.00 	 &	-5.48 	\\
090424	&	0.544 	&	49.47 	&	0.12 	&	300 	& a	& -1.33 	&	-2.37 	&	-4.78 	&	-5.26 	&	-2.31 	&	-3.35 	&	-5.27 	 &	-5.74 	\\
090812	&	2.452 	&	75.09 	&	1.85 	&	501 	& a	& -0.96 	&	-1.99 	&	-4.21 	&	-4.68 	&	-1.94 	&	-2.97 	&	-4.69 	 &	-5.17 	\\
090902B	&	1.8229 	&	21.9 	&	41.25 	&	120 	& a	& 0.66 	    &	-0.37 	&	-2.07 	&	-2.54 	&	-0.32 	&	-1.35 	&	-2.55 	 &	-3.03 	\\
090926A	&	2.1062 	&	20 	    &	29.35 	&	150 	& a	& 0.15 	    &	-0.89 	&	-2.49 	&	-2.96 	&	-0.83 	&	-1.87 	&	-2.97 	 &	-3.45 	\\
091024	&	1.092 	&	1020    &	0.06 	&	69  	& a	& 2.19 	    &	1.15 	&	-2.65 	&	-3.12 	&	1.21 	&	0.17 	&	 -3.13 	&	-3.61 	\\
091029	&	2.752 	&	39.18 	&	0.71 	&	221 	& a	& -1.35 	&	-2.39 	&	-4.23 	&	-4.71 	&	-2.33 	&	-3.37 	&	-4.72 	 &	-5.19 	\\
100621A	&	0.542 	&	63.6 	&	0.11 	&	52  	& a	& 0.39 	    &	-0.65 	&	-3.19 	&	-3.66 	&	-0.59 	&	-1.63 	&	-3.67 	 &	-4.15 	\\
100724B	&	1.00 	&	111.6 	&	0.98 	&	162 	& b	& 0.64 	    &	-0.40 	&	-3.09 	&	-3.57 	&	-0.34 	&	-1.38 	&	-3.58 	 &	-4.06 	\\
100728B	&	2.106 	&	12.1 	&	0.77 	&	373 	& a	& -2.70 	&	-3.74 	&	-5.08 	&	-5.55 	&	-3.68 	&	-4.72 	&	-5.56 	 &	-6.04 	\\
100906A	&	1.727 	&	114.4 	&	0.80 	&	369 	& a	& -0.43 	&	-1.47 	&	-4.02 	&	-4.50 	&	-1.41 	&	-2.45 	&	-4.51 	 &	-4.98 	\\
110205A	&	2.22 	&	257 	&	0.70 	&	177 	& a	& 0.75 	    &	-0.28 	&	-3.16 	&	-3.64 	&	-0.23 	&	-1.26 	&	-3.65 	 &	-4.12 	\\
110213A	&	1.46 	&	48 	    &	0.33 	&	223 	& a	& -1.10 	&	-2.14 	&	-4.30 	&	-4.78 	&	-2.08 	&	-3.12 	&	-4.79 	 &	-5.26 	\\
130504C	&	1.00 	&	74 	    &	0.58 	&	144 	& b	& 0.12 	    &	-0.91 	&	-3.40 	&	-3.87 	&	-0.86 	&	-1.89 	&	-3.88 	 &	-4.36 	\\
130518	&	2.49 	&	48 	    &	10.0 	&	257 	& b	& -0.07 	&	-1.11 	&	-3.09 	&	-3.57 	&	-1.05 	&	-2.09 	&	-3.58 	 &	-4.06 	\\
130821A	&	1.00 	&	84 	    &	0.69 	&	72  	& b	& 0.92 	    &	-0.11 	&	-2.67 	&	-3.14 	&	-0.06 	&	-1.09 	&	-3.15 	 &	-3.63 	\\
131108A	&	2.40 	&	19 	    &	8.52 	&	907 	& b	& -2.09 	&	-3.12 	&	-4.65 	&	-5.13 	&	-3.07 	&	-4.10 	&	-5.14 	 &	-5.61 	\\
131231A	&	0.642 	&	31 	    &	0.54 	&	192 	& b	& -0.79 	&	-1.83 	&	-3.97 	&	-4.45 	&	-1.77 	&	-2.81 	&	-4.46 	 &	-4.93 	\\
140206B	&	1.00 	&	120 	&	1.28 	&	254 	& b	& 0.43 	    &	-0.60 	&	-3.34 	&	-3.81 	&	-0.55 	&	-1.58 	&	-3.82 	 &	-4.30 	\\
141028A	&	2.332 	&	31.5 	&	6.11 	&	397 	& b	& -1.02 	&	-2.05 	&	-3.85 	&	-4.32 	&	-2.00 	&	-3.03 	&	-4.33 	 &	-4.81 	\\
\enddata
\tablenotetext{1}{References: (a) L\"{u} et al. (2012); (b) Tang et al. (2014).}
\end{deluxetable}

\clearpage

\begin{figure*}
\centering
\includegraphics[angle=0,scale=0.5]{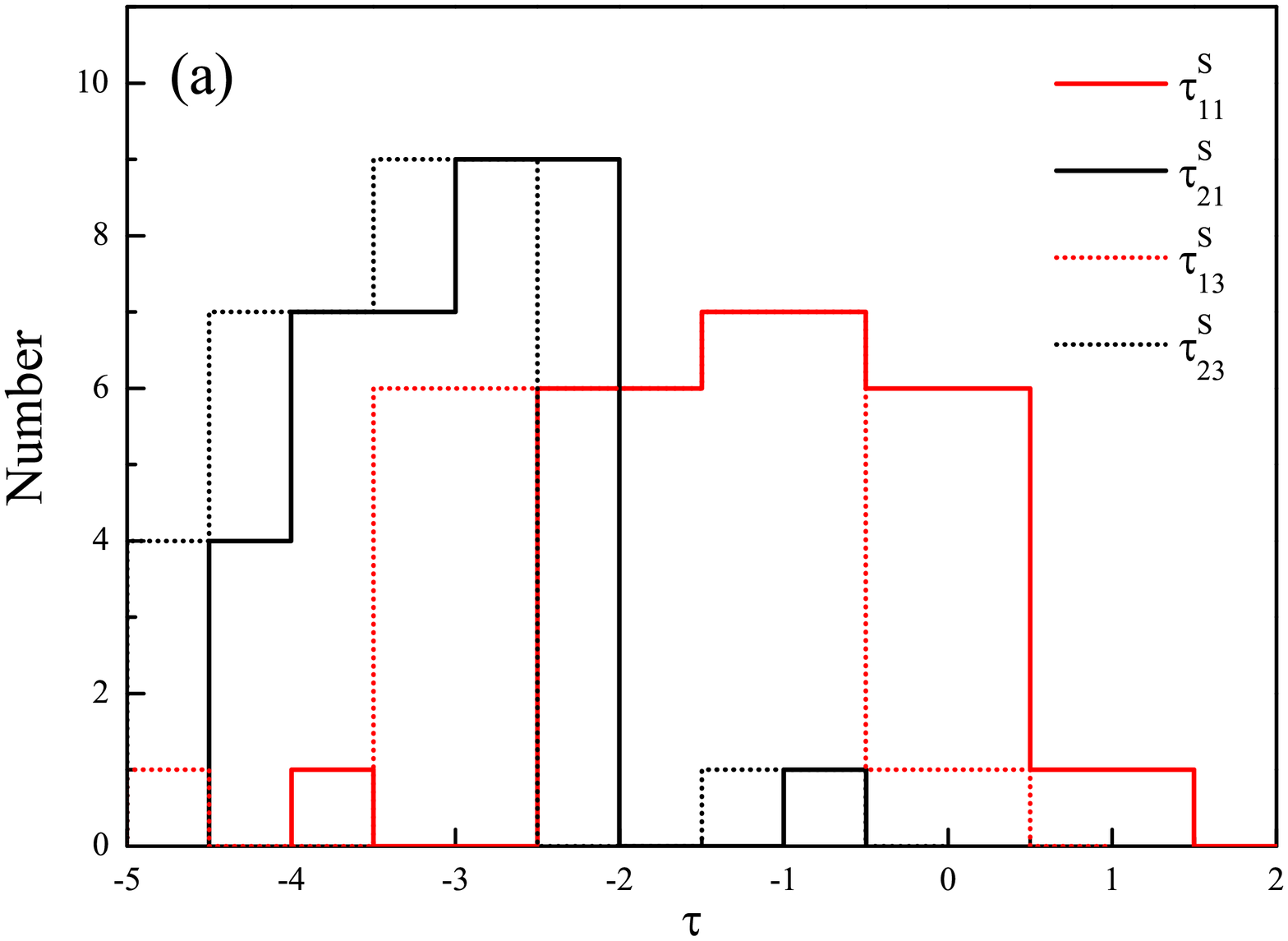}
\includegraphics[angle=0,scale=0.5]{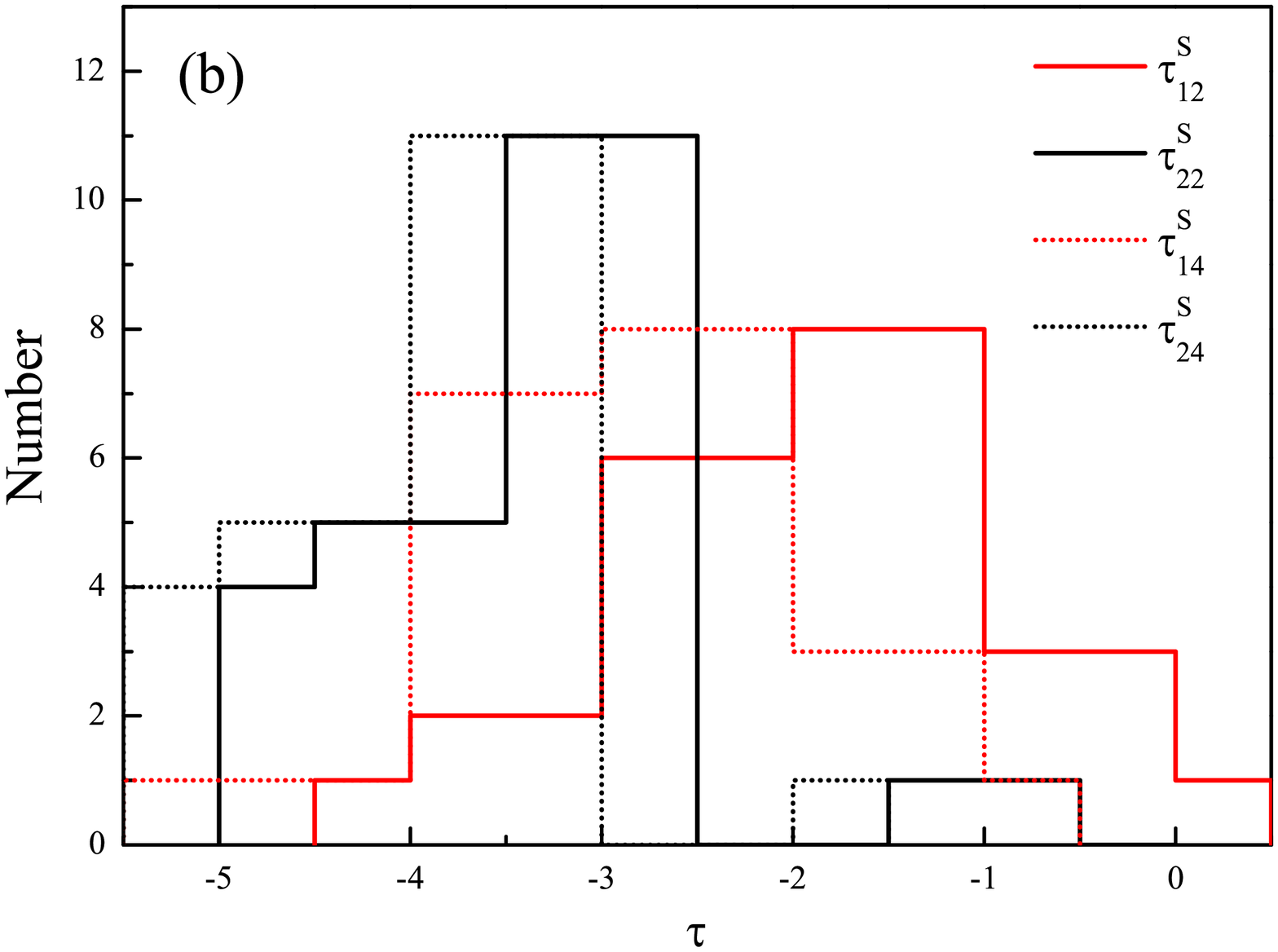}
\caption{Distributions of $\tau$ for SGRBs under the BZ mechanism or neutrino annihilation with the different BH spins and disk masses.}
\label{sample-figure1}
\end{figure*}

\clearpage

\begin{figure*}
\centering
\includegraphics[angle=0,scale=0.5]{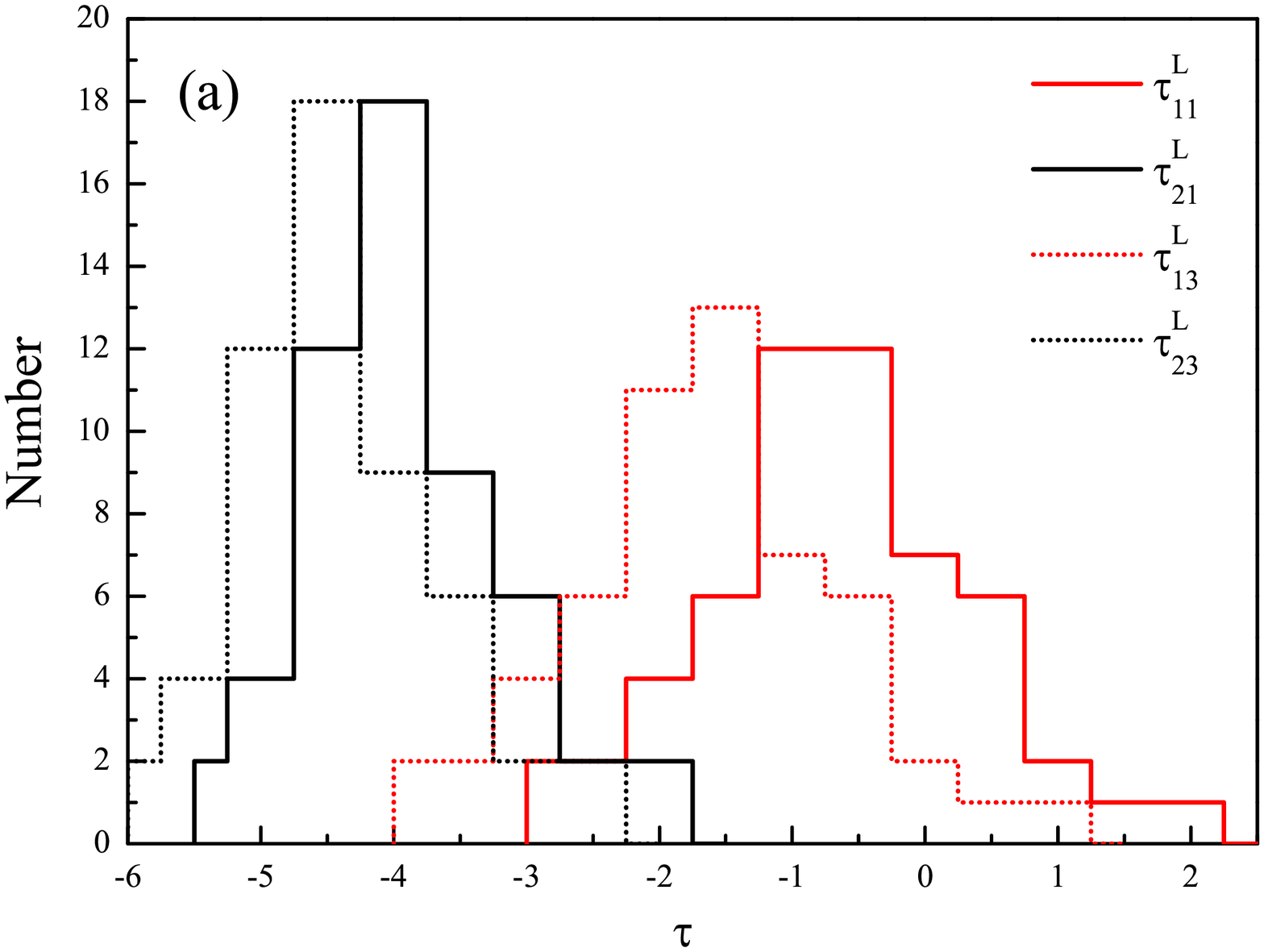}
\includegraphics[angle=0,scale=0.5]{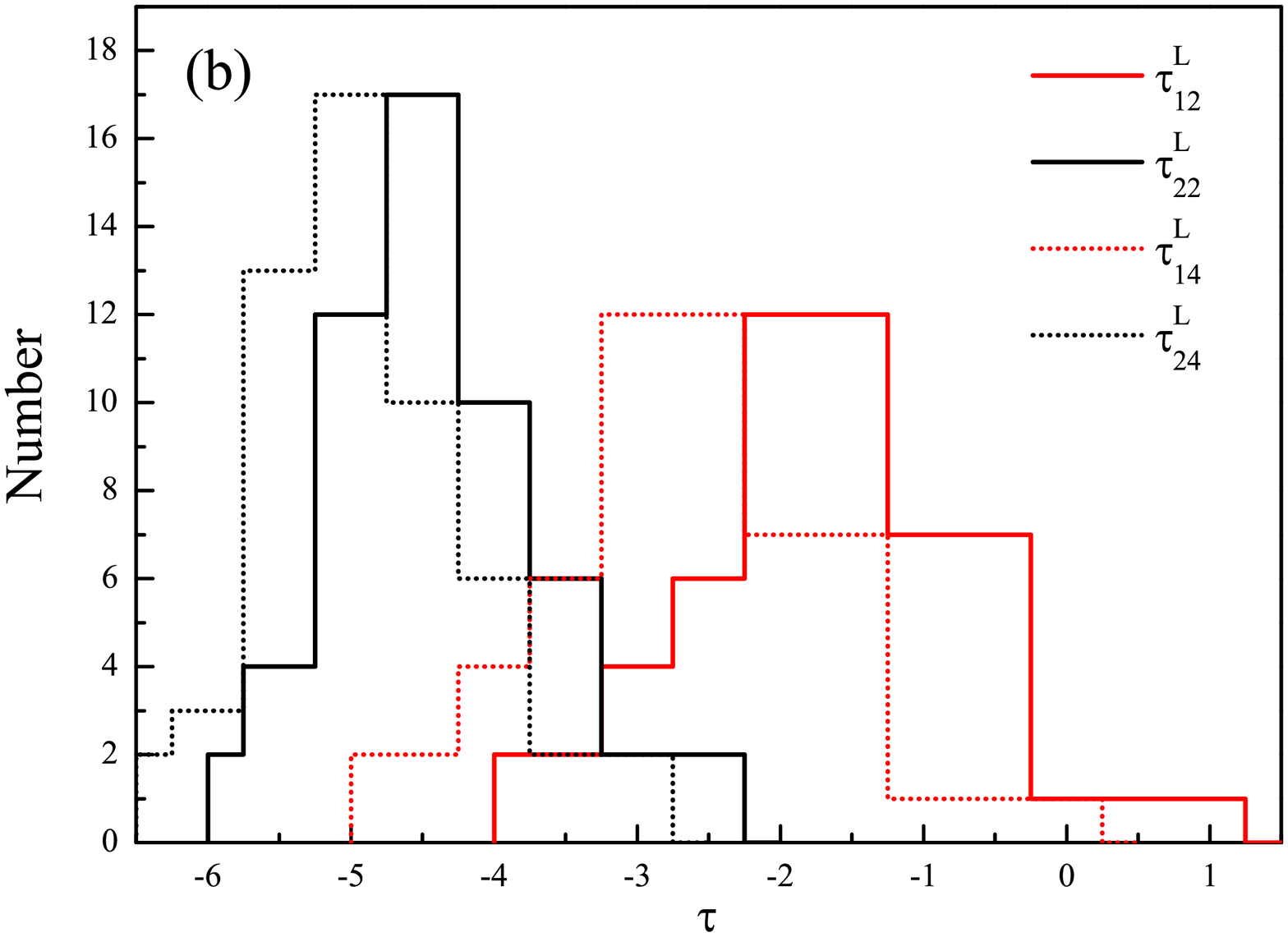}
\caption{Distributions of $\tau$ for LGRBs under the BZ mechanism or neutrino annihilation with the different BH spins and disk masses.}
\label{sample-figure2}
\end{figure*}

\clearpage

\end{document}